\documentclass[a4paper,12pt]{article}
\usepackage[utf8]{inputenc}
\usepackage[margin=2.5cm]{geometry}
\usepackage{graphicx}
\usepackage{amssymb,amsthm,amsmath,mathtools}
\usepackage{authblk}
\usepackage{algorithm}
\usepackage{algpseudocode}
\usepackage[title]{appendix}

\providecommand{\keywords}[1]
{
  \small	
  {\textit{Keywords: }} #1
}

\begin{document}

\title{Reconstructing initial pressure and speed of sound distributions simultaneously in photoacoustic tomography}

\author[a]{Miika Suhonen}
\author[b]{Felix Lucka}
\author[a]{Aki Pulkkinen}
\author[c]{Simon Arridge}
\author[d]{Ben Cox}
\author[a,c]{Tanja Tarvainen}

\affil[a]{\small \textit{University of Eastern Finland, Department of Technical Physics, Finland}}
\affil[b]{\textit{Centrum Wiskunde \& Informatica, The Netherlands}}
\affil[c]{\textit{University College London, Department of Computer Science, United Kingdom}}
\affil[d]{\textit{University College London, Department of Medical Physics and Biomedical
Engineering, United Kingdom}}

\date{}
\maketitle
            

\begin{abstract}
\noindent
Image reconstruction in photoacoustic tomography relies on an accurate knowledge of the speed of sound in the target. However, the speed of sound distribution is not generally known, which may result in artefacts in the reconstructed distribution of initial pressure. Therefore, reconstructing the speed of sound simultaneously with the initial pressure would be valuable for accurate imaging in photoacoustic tomography. Furthermore, the speed of sound distribution could provide additional valuable information about the imaged target. In this work, simultaneous reconstruction of initial pressure and speed of sound in photoacoustic tomography is studied. This inverse problem is known to be highly ill-posed. To overcome this, we study an approach where the ill-posedness is alleviated by utilising multiple photoacoustic data sets that are generated by different initial pressure distributions within the same imaged target. Then, these initial pressure distributions are reconstructed simultaneously with the speed of sound distribution. A methodology for solving this minimisation problem is formulated using a gradient-based iterative approach equipped with bound constraints and a multigrid approach. The methodology was evaluated with numerical simulations. Different approaches for generating multiple initial pressure distributions and their effect on the solution of the image reconstruction problem were studied. The results show that initial pressure and speed of sound can be simultaneously reconstructed from photoacoustic data. Furthermore, utilising multiple initial pressure distributions improves the reconstructions such that the locations of initial pressure and speed of sound inhomogeneities can be better distinguished and image artifacts are reduced.
\end{abstract}

\vspace{1cm}

\vspace{2cm}

\section{Introduction}
\label{sec:Introduction}

Photoacoustic tomography (PAT) is a biomedical imaging modality developed during recent decades \cite{Beard_2011, Li_2009}. In photoacoustic tomography, generation of acoustic pressure waves by the photoacoustic effect is triggered by an external excitation of a near-infrared light. These pressure waves propagate inside the target, and are measured on its boundary using ultrasound sensors. Then, an image of the target is reconstructed from these measurements. PAT is a hybrid imaging modality that combines the unique contrast of optical imaging and the high resolution of ultrasound. As a result, it has various potential applications in medical and biomedical imaging, such as cancer detection and monitoring, small animal imaging, and vascular imaging \cite{Attia_2019_Photoacoustics, Gu_2022_BTM, Li_2021_BMEF}. 

In the image reconstruction, i.e. inverse problem, of PAT, the initial pressure distribution is reconstructed. Different image reconstruction methods for PAT have been developed, such as analytical methods, filtered back-projection,  time reversal, model-based variational methods, the Bayesian approach, and machine learning methods \cite{Ben_2022, Hauptmann_2020_JBO, Kuchment_2011_Springer, Poudel_2019, Tick_2016, Xu_2005}. Many of the methods assume that the propagation of the acoustic waves occurs in a homogeneous medium, i.e. assuming a constant speed of sound. However, this assumption is rarely valid in real biological tissues. If the heterogeneous speed of sound is not taken into account, reconstructed initial pressure distributions can suffer from artefacts and aberrations, see e.g. Ref. \cite{Wang_2020_JIOHS} and references therein. 

Different approaches for compensating the unknown speed of sound have been studied. For example, the "best possible" constant speed of sound has been selected by performing PAT image reconstructions using different speed of sound values and then selecting the one that maximises some image quality metrics \cite{Mandal_2014_Photoacoustics, Treeby_2011_JBO}. However, reconstructed images can still suffer from artefacts in targets containing heterogeneous speed of sound. Alternatively, other approaches, such as deep learning \cite{Jeon_2021_TIP, Poimala_2024_Photoacoustics} and Bayesian approximation error modelling \cite{Tick_2020_BPE}, have been used to compensate errors due to using constant speed of sound in the image reconstruction. These methods rely on the training datasets and models, that need to be accurate in terms of good generalisation. One approach to overcome the problem of unknown speed of sound has been to utilise ultrasound measurements to determine the speed of sound. This has been implemented by first reconstructing the speed of sound distribution from ultrasound measurements, and then reconstructing the initial pressure from photoacoustic data utilising the predefined speed of sound in the reconstruction process \cite{Dantuma_2023_arXiv, Jose_2012_MedPhys, Mercep_2019_LSA, Ranjbaran_2024_TMI, Xia_2013_OptLet}. Alternatively, in Ref. \cite{Matthews_2017_IP} an approach where the speed of sound and initial pressure were simultaneously reconstructed using both full wave-field ultrasound tomography data and photoacoustic data was proposed. Reconstruction of speed of sound from additional data poses more requirements for the measurement system, and also increases the computational complexity of the image reconstruction process.  

An ideal approach, as it would require no additional hardware, to overcome challenges due to unknown speed of sound would be to reconstruct the speed of sound simultaneously with the initial pressure from photoacoustic data alone. In addition, when reconstructing the speed of sound simultaneously with the initial pressure, additional quantitative information about the target could be gained, since the reconstructed speed of sound can provide interesting knowledge about the properties of the target tissue. For example, the aim in ultrasound computed tomography is to reconstruct a spatially varying speed of sound and utilise that information in medical or biomedical purposes \cite{Duric_2018_Intech, Li_2009_UMB}. In Ref. \cite{Stefanov_2013_IPI}, the linearised problem of reconstructing the initial pressure and speed of sound simultaneously was shown to be unstable, suggesting the instability of the nonlinear problem. On the other hand, simultaneous reconstruction of spatially varying initial pressure and a constant speed of sound has been shown to have a unique solution \cite{Liu_2015_IP}. In the case where both of these parameters are spatially varying, a unique solution for the reconstruction problem can be found only when additional assumptions are made \cite{Acosta_2019_IP, Kian_2023_arXiv, Knox_2020_IP}. In Ref. \cite{Huang_2016_TCI}, the instability of this problem was demonstrated with numerical simulations. These various studies, such as in Refs. \cite{Stefanov_2013_IPI, Huang_2016_TCI}, showing the ill-posedness of the problem, indicate that simultaneous reconstruction of initial pressure and speed of sound is highly difficult in practice without further prior knowledge.

The ill-posedness of the inverse problem can be mitigated using different approaches. For example, in the variational approach, regularisation can be included to alleviate the ill-posedness of the image reconstruction problem \cite{Scherzer_2009_Springer}. Similarly, in the Bayesian framework \cite{Kaipio_2005}, prior information of the imaged target can be incorporated, leading to improved condition of the inverse problem. These approaches have been utilised in PAT in simultaneous reconstruction of the initial pressure and speed of sound. In Ref. \cite{Jeong_2025_Photoacoustics}, the effect of adding tight constraints, based on prior knowledge about the imaged target, to the optimisation problem was studied indicating that the ill-posedness of the problem can be mitigated when more prior knowledge is included. In Refs. \cite{Matthews_2018_JIS, Zhang_2008_ProcSPIE}, a model-based variational reconstruction of initial pressure and speed of sound, when the speed of sound distribution was re-parameterised to a coarse discretisation, was proposed. This enabled reconstructing both of these parameters in a more stable manner. Furthermore, in Refs. \cite{Cai_2019_BOE, Deng_2022_JBO, Zhang_2006_ProcSPIE} a so-called feature coupling approach was proposed, where the detector arrays were divided into subsets and different initial pressure distributions were reconstructed from each subset. Then, the speed of sound, in coarse discretisation, was reconstructed simultaneously with the initial pressure by maximising the correlation of the different photoacoustic reconstructions. However, these methods still require prior knowledge on the locations of different tissues to perform the coarse discretisation accurately. 

In this work, we propose an approach for PAT where multiple initial pressure distributions are reconstructed simultaneously with the speed of sound distribution. With this approach, the number of unknown parameters grows compared to a situation of a single initial pressure distribution. However, more data containing information about the unknown speed of sound distribution is gained. This alleviates the ill-posedness of the problem, and thus facilitates numerical solving of the related minimisation problem. In practise, these different initial pressure distributions could be generated using various approaches. For example, imaged target can be illuminated from different directions or by using spatially modulated illuminations patterns. This approach is utilised in quantitative PAT to overcome the non-uniqueness of simultaneous reconstruction of optical absorption and scattering coefficients, see Refs. \cite{Bal_2011, Tarvainen_2012, Zemp_2010} for directional and Ref. \cite{Pulkkinen_2015_JBO} for spatially modulated illuminations. In addition, illuminations from different directions have been utilised in simultaneous reconstruction of optical absorption and the speed of sound distributions in quantitative PAT \cite{Ding_2015_IP}. Another possibility to generate different initial pressure distributions would be to use a single illumination pattern but different wavelengths of light. Furthermore, an approach where an additional optical absorber is placed to the imaged target, such as on its boundary, could be utilised, for example, similarly as in Ref. \cite{Manohar_2007_APL} for laser-induced ultrasound tomography. The position of this absorber could be varied, and thus the external photoacoustic source would lie in different position, generating different initial pressure distribution that would result into acoustic pressure waves from different positions. In this work, to study the proposed approach, different initial pressure distributions are generated using three approaches: by using illuminations from different directions, by using different wavelengths of light, and by adding additional absorbers to the target.

The rest of the paper is organised as follows. In Sec. \ref{sec:methodology}, the forward modelling of photoacoustic tomography and simultaneous reconstruction of the initial pressure and speed of sound are presented. In Sec. \ref{sec:Optimisation}, optimisation strategies used in the image reconstruction are presented. In Sec. \ref{sec:Simulations}, the numerical simulations to evaluate the methodology are described and their results are shown. Furthermore, the results are discussed and conclusions are given in Sec. \ref{sec:discussion}.

\section{Methodology}
\label{sec:methodology}

\subsection{Forward model}
\label{sec:forward_problem}

Propagation of acoustic waves generated by an initial pressure distribution is known as the initial value problem in acoustics. This initial value problem for the wave equation in an acoustically heterogeneous non-absorbing medium is given by
\begin{equation}
    \begin{cases}
    &\dfrac{1}{c^2(r)}\dfrac{\partial^2}{\partial t^2}p(r,t) - \nabla^2 p(r,t) = 0, \quad r\in \mathbb{R}^d, \: t \in [0,T] \\
    &p(r,t=0) = p_0(r), \\
    &\dfrac{\partial}{\partial t}p(r,t=0) = 0,
    \end{cases}
    \label{eq:WaveEquation}
\end{equation}
where $r$ is the position, $d$ is the dimension, $p_0(r)$ is the initial pressure distribution, $p(r,t)$ is the acoustic pressure at a time $t$, and $c(r)$ is the speed of sound \cite{Cox_2007_JASA}. In this work, we approximate the solution of the wave equation (\ref{eq:WaveEquation}) using a \textit{k}-space pseudo-spectral method implemented in the k-Wave MATLAB toolbox \cite{Treeby_2010_kwave}.

\subsection{Image reconstruction}
\label{sec:inverse_problem}

In this work, data is considered to consist of multiple photoacoustic datasets that are created by different initial pressure distributions in the imaged target. Then, the aim of the inverse problem is to simultaneously reconstruct these different initial pressure distributions and the speed of sound distribution.

Let us consider a situation where $i = 1,...,I$ different initial pressure distributions $p_{0}^{i}(r)$ are generated inside the imaged target. Let us use a notation $p_{0}^{i} = \left( p_{0,1}^{i},...,p_{0,N}^{i} \right) \in \mathbb{R}^{N}$ to describe an initial pressure distribution represented in a spatial discretisation, where $N$ is the number of spatial discretisation points. Similarly, the discretised speed of sound is $c = \left( c_{1},...,c_{N} \right) \in \mathbb{R}^{N}$. The photoacoustic (measurement) data $y^{i} \in \mathbb{R}^{M}$, that is corrupted with measurement noise $e^{i} \in \mathbb{R}^{M}$, is generated by the initial pressure $p_{0}^{i}$. This data is sampled at a discrete set of sensor locations with a finite number of time steps. Further, let us denote the unknown multiple initial pressure distributions and speed of sound as $x = \left( p_{0}^{1},...,p_{0}^{I}, c \right)^{\top} \in \mathbb{R}^{(I+1)N}$, full (measurement) data as $y = \left( y^{1},...,y^{I}\right)^{\top} \in \mathbb{R}^{IM}$ and noise $e = \left( e^{1},...,e^{I} \right)^{\top} \in \mathbb{R}^{IM}$. The observation model with an additive noise can be written as
\begin{equation}
    y = f(x) + e,
    \label{eq:ObservationModel}
\end{equation}
where $f(x) = \left( f^{1}(x^{1}),...,f^{I}(x^{I}) \right)^{\top}, \: f^{i}: \mathbb{R}^{2N} \rightarrow \mathbb{R}^{M}$ is the discretised forward operator that maps unknown parameters $x^{i} = \left( p_{0}^{i}, c\right)^{\top} \in \mathbb{R}^{2N}$ to the data $y^{i}$.

The solution to the inverse problem can be calculated by solving an optimisation problem
\begin{align}
    \hat{x} = \underset{x}{\arg \min}\left\{ \: \varepsilon(x) \: \right\},
    \label{eq:Optimisation_Problem}
\end{align}
where $\varepsilon(x)$ is the objective function to be minimized. In this work, the objective function is of the form
\begin{align}
    \varepsilon(x) =  \sum_{i=1}^{I} \frac{1}{2}\left\Vert  L_{e^{i}} \left( y^{i} - f^{i}(x^{i}) - \eta_{e^{i}} \right) \right\Vert^2 + \sum_{i=1}^{I} \frac{1}{2}\left\Vert L_{p_{0}^{i}} \left( p_{0}^{i} - \eta_{p_{0}^{i}} \right) \right\Vert^2 + \frac{1}{2}\left\Vert L_{c} \left( c - \eta_{c} \right)\right\Vert^2,
    \label{eq:Minimised_functional}
\end{align}
where the first term is the data likelihood, where $\eta_{e^{i}}$ is the mean of the noise and $L_{e^{i}}$ is a weighting matrix, that in a Bayesian framework is the Cholesky decomposition of the inverse of the noise covariance matrix $L_{e^{i}}^{\top}L_{e^{i}}^{} = \Gamma_{e^{i}}^{-1}$ \cite{Kaipio_2005, Tarantola_2005, Tick_2016}. Further, the last two (regularising) terms are the priors for the initial pressure and the speed of sound, where $\eta_{p_{0}^{i}}$ and $\eta_{c}$ are the means of the initial pressure and speed of sound, respectively, and $L_{p_{0}^{i}}^{\top}L_{p_{0}^{i}}^{} = \Gamma_{p_{0}^{i}}^{-1}$ and $L_{c}^{\top}L_{c}^{} = \Gamma_{c}^{-1}$ are the Cholesky decompositions, where $\Gamma_{p_{0}^{i}}$ and $\Gamma_{c}$ are the covariance matrices of the initial pressure and speed of sound.


\section{Optimisation}
\label{sec:Optimisation}

Optimisation problem (\ref{eq:Optimisation_Problem}) can be solved using iterative methods. In this work, the limited memory Broyden–Fletcher–Goldfarb–Shanno (L-BFGS) method is used \cite{ Gao_2015, Nocedal_2006}. L-BFGS requires only the gradients of the objective functional to be calculated, and thus no second order derivatives are needed. Evaluation of the gradients can be implemented with the adjoint-state method.

Let us define the adjoint problem of the wave equation (\ref{eq:WaveEquation}) as
\begin{equation}
    \begin{cases}
    &\dfrac{1}{c^2(r)}\dfrac{\partial^2}{\partial t^2}q(r,t) - \nabla^2 q(r,t) = S(r,t), \quad r\in \mathbb{R}^d, \: t \in [0,T] \\
    &q(r,0) = 0, \\
    &\dfrac{\partial}{\partial t}q(r,0) = 0,
    \end{cases}
    \label{eq:AdjointEquation}
\end{equation}
where $q(r,t)$ is the adjoint acoustic pressure (the solution of the adjoint equation) and $S(r,t) = p(r,T-t)w(r,t)$ is the time-varying source term where $w(r,t)$ is a window function mapping the pressure field accessible to the sensors \cite{Arridge_2016_IP}. Considering the discretised parameters and models defined in Sec. \ref{sec:inverse_problem}, the adjoint field $q^{i} = \left( q_{1}^{i},...,q_{N}^{i} \right)^{\top} \in \mathbb{R}^{NT}$ can be numerically approximated by setting a (discrete) source to the discretised adjoint equation (\ref{eq:AdjointEquation}) as
\begin{align}
    S^{i} = \left( L_{e^{i}}^{\top}L_{e^{i}}^{} \right) \left( y^{i} - f^{i}(x^{i}) \right),
\end{align}
where term $L_{e^{i}}^{\top}L_{e^{i}}^{}$ is acting as a weight to the residual $\left( y^{i} - f^{i}(x^{i}) \right)$. The functional gradient for the initial pressure, i.e. the gradient of Eq. (\ref{eq:Minimised_functional}) with respect to initial pressure $p_{0}^{i}$, can be calculated with the adjoint-state method as
\begin{align}
    \dfrac{\partial \varepsilon(x)}{\partial p_{0}^{i}} &= q^{i}(r,T) + L_{p_{0}^{i}}^{\top}L_{p_{0}^{i}}^{} \left( p_{0}^{i} - \eta_{p_{0}^{i}} \right),
    \label{eq:gradient_p0}
\end{align}
where $q^{i}(r,T)$ is the solution of the adjoint equation (\ref{eq:AdjointEquation}) at discretised position $r$ at time instance $t=T$. In addition, the functional gradient with respect to speed of sound $c$ can be written as \cite{Bunks_1995_Geophysics, Norton_1999_JASA}
\begin{align}
    \dfrac{\partial \varepsilon(x)}{\partial c} &= \sum_{i=1}^{I} \int_{0}^{T} \left(-\dfrac{2}{c^3}\dfrac{\partial^2p^{i}(r,t)}{\partial t^2} \right)^{\top} q^{i}(r,T-t) \: dt + L_{c}^{\top}L_{c}^{} \left( c - \eta_{c} \right).
    \label{eq:gradient_sos}
\end{align}

Let us denote the gradient of the objective function (\ref{eq:Minimised_functional}) as 
\begin{align}
    \frac{\partial \varepsilon(x)}{\partial x} = \left( \frac{\partial \varepsilon(x)}{\partial p_{0}^{1}},...,\frac{\partial \varepsilon(x)}{\partial p_{0}^{I}}, \frac{\partial \varepsilon(x)}{\partial c} \right)^{\top}.
\end{align} 
On iteration $k$ of the L-BFGS algorithm, the parameter vector $x$ is updated as follows
\begin{align}
    x_{k+1} = \mathcal{P}\left( x_{k} - \alpha_{k}H_{k}\dfrac{\partial \varepsilon(x_{k})}{\partial x_{k}} \right),
        \label{eq:BFGS_iterations}
\end{align}
where $H_{k}$ is the inverse Hessian approximation, $\mathcal{P}$ is the projector operator for bound constraints, and $\alpha_{k}$ is the step length parameter. In L-BFGS, the approximation of the inverse Hessian $H_{k}$ is constructed by using $l$ previous gradient and parameter vector values \cite{Nocedal_2006}. This can be performed by updating $H_k$ with a formula
\begin{align*}
    H_{k} =& \left( V_{k-1}^{\top} \cdots V_{k-l}^{\top} \right) H_{k}^{0} \left( V_{k-l}^{} \cdots V_{k-1}^{} \right) \\
    &+ \rho_{k-l} \left( V_{k-1}^{\top} \cdots V_{k-l+1}^{\top} \right) \varkappa_{k-l}^{}\varkappa_{k-l}^{\top} \left( V_{k-l+1}^{} \cdots V_{k-1}^{} \right) \\
    &+ \rho_{k-l+1} \left( V_{k-1}^{\top} \cdots V_{k-l+2}^{\top} \right) \varkappa_{k-l+1}^{}\varkappa_{k-l+1}^{\top} \left( V_{k-l+2}^{} \cdots V_{k-1} ^{}\right) \\
    &+ \cdots \\
    &+ \rho_{k}^{}\varkappa_{k}^{}\varkappa_{k}^{\top},
\end{align*}
where $\varkappa_{k} = x_{k} - x_{k-1}$, $\rho_{k} = \frac{1}{\varphi_{k}^{\top}\varkappa_{k}^{}}$, $V_{k} = \mathbb{I} - \rho_{k}^{}\varphi_{k}^{}\varkappa_{k}^{\top}$, $\varphi_{k} = \frac{\partial \varepsilon(x_{k})}{\partial x_{k}} - \frac{\partial \varepsilon(x_{k-1})}{\partial x_{k-1}}$ and $\mathbb{I}$ is an identity matrix. In the equation, initial inverse Hessian approximation $H_{k}^{0}$ is chosen as $H_{k}^{0} = \frac{\varkappa_{k}^{\top}\varphi_{k}^{}}{\varphi_{k}^{\top}\varphi_{k}^{}}\mathbb{I}$. Implementation of the L-BFGS update direction can be performed with a two-loop recursion, described in Algorithm \ref{alg:L_BFGS} \cite{Nocedal_2006}.
\begin{algorithm}[tb]
\caption{L-BFGS update with two-loop recursion}\label{alg:L_BFGS}
\begin{algorithmic}
\State $\phi \gets \frac{\partial \varepsilon(x_{k})}{\partial x_{k}}$
\For{$j =k-1,k-2,...,k-l$}
    \State $\lambda_j \gets \rho_j^{} \varkappa_j^{\top }q$
    \State $\phi \gets \phi - \lambda_j \varphi_i$
\EndFor
\State $\psi \gets H_{k}^{0} \phi$
\For{$j = k-l,k-l+1,...,k-1$}
    \State $\beta \gets \rho_j^{} \varphi_j^{\top} \psi $
    \State $\psi \gets \psi - \varkappa_j(\lambda_j - \beta)$
\EndFor
\State $H_{k}\frac{\partial \varepsilon(x_{k})}{\partial x_{k}} \gets \psi$
\end{algorithmic}
\end{algorithm}
The operator $\mathcal{P}(z)$ to implement the bound constraints can be written as
\begin{align}
\mathcal{P}(z) = \begin{cases}
    L, \quad &z \leq L \\
    z, \quad &L < z < U \\
    U, \quad &U \leq z, 
    \end{cases}
    \label{eq:bound_constraints}
\end{align}
where $L$ and $U$ are the lower and upper bounds. Furthermore, the step length is evaluated on each iteration with a backtracking line-search \cite{Nocedal_2006}.

\subsection{Improving convergence}
\label{sec:multigrids}

In seismic and ultrasound imaging, optimisation problem (\ref{eq:Optimisation_Problem})-(\ref{eq:Minimised_functional}) is non-convex. This means that, when iteratively solving the inverse problem, it is possible to get stuck in a local minimum without finding the global minimum. This issue is often referred to as "a cycle skipping problem", as it typically leads to a phenomenon where cycles of the sinusoidal simulated data traces are matched to incorrect ones in the (measured) time series data \cite{Virieux_2009_Geophysics}. Similar problem can occur in PAT when the speed of sound is simultaneously reconstructed with the initial pressure. To avoid the cycle skipping problem, one can try to reduce the high frequencies in the early stage of the iteration algorithm and increase the frequency content as iteration proceeds. This can be achieved, for example, by modifying the order of spatial discretisation during the iteration. In addition, different approaches to tackle the cycle skipping problem have been proposed, for example, low-pass filtering \cite{Bunks_1995_Geophysics, Bernard_2017_PMB} and modifying the objective function \cite{Warner_2016_GeoPhys}. 

In this work, we utilise the multigrid approach, where different discretisations \cite{Bunks_1995_Geophysics, Lucka_2022_IP} are used during the iterative image reconstruction process to incorporate frequency content gradually to avoid the cycle skipping problem. The image reconstruction process is started with a coarse discretisation and the discretisation order is increased during the iterations. In every discretisation, the iterative image reconstruction process is continued until the solution converges. The reconstructed parameters are then linearly interpolated to a finer discretisation and the interpolated values for the parameter distribution are used as an initial guess for the iteration process in the finer discretisation. This is repeated until the problem has converged in the finest discretisation. The approach does not only help to avoid getting stuck in local minimums, but it also reduces the total computation time, since iterations are started with coarser discretisations that are computationally less expensive  \cite{Javaherian_2017_TMI, Li_2015_BOE}.

\section{Simulations}
\label{sec:Simulations}

To evaluate the proposed methodology, numerical simulations in 2D were performed. These were carried out in a rectangular $20 \: \mathrm{mm} \times 20 \: \mathrm{mm}$ domain. In all simulations, $236$ acoustic point-like sensors were evenly placed in a distance of $0.2 \: \mathrm{mm}$ from the boundary covering all sides of the domain. This setup simulates a planar detection geometry such as Fabry-P\'{e}rot scanner, see e.g. Refs. \cite{Ellwood_2017_JBO, Huynh_2024_NatureBE, Zhang_2008_AppliedOptics}. We studied the simultaneous reconstruction of the initial pressure and the speed of sound with the proposed approach where multiple different initial pressure distributions were utilised. The approach was compared against (conventional) approach, where only one initial pressure distribution and the speed of sound distribution were reconstructed simultaneously.

\subsection{Simple phantom}
\label{sec:SimplePhantom}

\subsubsection{Phantom and data simulation}
\label{sec:SimplePhantom_TargetAndDatasim}

In the first set of simulations, a simplified numerical phantom consisting of different initial pressure and speed of distributions was used to study the proposed methodology for simultaneous reconstruction of these parameters. 

First, a numerical phantom consisting of circular inclusions, was studied (see Fig. \ref{fig:JR_simplecircular} in Sec. \ref{sec:SimplePhantom_Results}). The speed of sound values were of low-contrast, mimicking typical values of soft biological tissue, with the target consisting of nine circular inclusions with $1580 \: \mathrm{m/s}$ in a constant background with $1430 \: \mathrm{m/s}$. For the proposed approach, four different initial pressure distributions $p_{0}^{1}$, $p_{0}^{2}$, $p_{0}^{3}$ and $p_{0}^{4}$ were simulated. These consisted of a circular inclusions in a zero background. The initial pressure in the inclusions was set arbitrarily to $10 \: \mathrm{Pa}$, but due to linearity of the system, this could be scaled. In the reference approach, the initial pressure distribution $p_{0}^{1}$ was used.

In the second simulation, a similar numerical phantom as previously was used but the speed of sound distribution included also a water bath mimicking layer, which is typical in many applications of photoacoustic tomography (Fig. \ref{fig:JR_simplecircular_water} in Sec. \ref{sec:SimplePhantom_Results}). The speed of sound values of the background, circular inclusions and water bath was set to $1430 \: \mathrm{m/s}$, $1580 \: \mathrm{m/s}$ and $1482 \: \mathrm{m/s}$, respectively. The same four different initial pressure distributions $p_{0}^{1}$, $p_{0}^{2}$, $p_{0}^{3}$ and $p_{0}^{4}$ as in the first simulation were used. The initial pressure distribution $p_{0}^{1}$ was used in the reference approach. 

In the third simulation, the speed of sound distribution consisted, in addition to the circular inclusions, of a rectangular inclusion with a higher contrast to the background, mimicking values typical of a bone-like tissue (Fig. \ref{fig:JR_simplecircular_bone} in Sec. \ref{sec:SimplePhantom_Results}). The speed of sound in the background, circular inclusions and bone mimicking inclusion were set to $1430 \: \mathrm{m/s}$, $1580 \: \mathrm{m/s}$ and $2500 \: \mathrm{m/s}$, respectively. Four different initial pressure distributions $p_{0}^{1}$, $p_{0}^{2}$, $p_{0}^{3}$ and $p_{0}^{4}$ were used also in this simulation. In the reference approach, the initial pressure distribution $p_{0}^{1}$ was used, similarly as in the previous simulations.

To simulate the photoacoustic data from an initial pressure, the target domain was discretised into $335 \times 335$ pixel grid with a pixel size of $63.7 \: \mathrm{\mu m}$. A small layer of $10$ pixels, where the acoustic sensors were located, was added outside of the target domain. In addition, a perfectly matched layer (PML) of $20$ pixels was added to each side (outside of the target and sensor domains) to reduce reflections on the boundary. The size of the PML-layer was chosen such that the overall grid size would have the smallest prime factors to reduce computational time with k-Wave \cite{Treeby_2010_kwave}. In the temporal discretisation, $1747$ time points with a time step of $12.1 \: \mathrm{ns}$  was used. These choices enabled the acoustic simulations with a peak frequency support of $11.2 \: \mathrm{MHz}$. Discretisation parameters used in the data simulation are listed in Table. \ref{tab:GridParameters}. In all simulations, additive Gaussian distributed noise with $1 \%$ of peak-to-peak amplitude of the corresponding simulated photoacoustic data, was added to the data.

\begin{table*}[tb!]
\renewcommand{\arraystretch}{1.2}
\centering
\caption{Discretisation parameters of the data simulation grid and image reconstruction (multi) grids of different orders: pixel size $\Delta z \: \mathrm{(\mu m)}$, the number of pixels $N_z$ in the target domain, the number of pixels to place the acoustic sensors $N_z^{\mathrm{S}}$ added outside the target domain on each side, the number of pixels in the PML $N_z^{\mathrm{PML}}$ added outside the target and sensor domains on each side, time discretization $\Delta t \: \mathrm{(ns)}$, number of time points $N_t$, and peak frequency $f_p  \: \mathrm{(MHz)}$ supported by the grid. Multigrid discretisations 1-4 were used for simple target and 2-5 were used for tissue-mimicking target.}
\begin{tabular}{l c c c c c c c}
\hline
 Grid & $\Delta z$ & $N_z$ & $\: N_z^{\mathrm{S}}$ & $\: N_z^{\mathrm{PML}}$ &  $\Delta t$ & $N_t$ & $f_p$ \\ \hline
 Data simulation & 63.7 & $335$ & $10$ & $20$ & 12.1& 1747& 11.2   \\
 Multigrid discretisation $1$  & 526.3 &  39&  2&  19& 99.9 & 212 & 1.4   \\ 
 Multigrid discretisation $2$  & 370.4 & 55 & 4 & 31 & 70.4 & 301 & 2.0   \\ 
 Multigrid discretisation $3$  & 227.3 &  89&  4&  14& 43.2 & 490 & 3.2   \\
 Multigrid discretisation $4$  & 156.3 &  129&  6&  17& 29.7 & 712 &  4.1  \\
 Multigrid discretisation $5$  & 101.0 & 199 & 8 & 14 & 19.2 & 1101 & 6.4   \\ \hline
\end{tabular}
\label{tab:GridParameters}
\end{table*}

\subsubsection{Image reconstruction}
\label{sec:SimplePhantom_ImageRec}

Image reconstruction was performed using a different discretisation compared to data simulation to avoid the inverse crime. In all situations, the multigrid approach (Sec. \ref{sec:multigrids}) was used, and the computational domain was discretised using four different grids, starting from the coarsest grid and gradually moving to more dense grids. The simulated noisy measurement data was interpolated to each temporal grid used in the image reconstruction. Details of the discretisations that were used in the image reconstruction are given in Table \ref{tab:GridParameters}. 

As the prior model, the Ornstein-Uhlenbeck prior was used \cite{Tick_2016, Rasmussen_2006, Suhonen_2024_JOSAA}. In the Ornstein-Uhlenbeck prior, the covariance matrix for each distinctive parameter distribution $z$ (different initial pressures and the speed of sound) can be presented as $\Gamma_{z}^{} = \sigma_{z}^2 \Xi$, where $\sigma_z$ is the standard deviation and the unit covariance matrix is defined as $\Xi_{ij} = \mathrm{exp}\left( -\frac{||r_i - r_j||}{\tau} \right)$. Here, $r_i$ and $r_j$ are the pixel locations and characteristic length scale $\tau$ controls the amount of spatial smoothness of adjacent pixels. The prior mean $\eta_z$ for each parameter was chosen as the median value of the interval of the true parameter distribution in the case of targets with low contrast speed of sound. For the target, where the speed of sound had a higher contrast, prior mean for speed of sound was chosen as $1482 \: \mathrm{m/s}$ i.e. value between the background and circular inclusions of the true parameter distribution. Standard deviations $\sigma_z$ were chosen so that $\sigma_z = 1/4\left( \max(z_{\mathrm{true}}) - \min(z_{\mathrm{true}})  \right)$. With these choices, it is assumed that $95\%$ of the reconstructed parameter values would lie within the values of the true parameter range, except for the speed of sound distribution in high contrast simulations, where most of the parameter values were expected to lie between the values of true background and circular inclusions. Characteristic length scale value $\tau = 1.5 \: \mathrm{mm}$ was used. Noise was modelled as additive with zero mean and a standard deviation corresponding to 1\% of the peak-to-peak amplitude of the simulated noisy signal for every dataset corresponding to different initial pressures. 

Throughout the work, memory value $l = 20$ in the L-BFGS method was used. Initial guess $x_0$ for the reconstructed parameters was chosen such that the initial pressure was zero and speed of sound was the mean of the prior. In addition, lower bound $L_{p_0} = 0$ for initial pressures was used, with no upper bound throughout the work. For the speed of sound, bounds $L_c = 1300 \: \mathrm{m/s}$ and $U_c= 1800 \: \mathrm{m/s}$ were used, assuming the values of speed of sound lies within realistic interval for soft tissue targets \cite{Szabo_2014}, except in simulations with a high speed of sound contrast, the upper bound $U_c= 3000 \: \mathrm{m/s}$ was used.

To evaluate the reconstructed images quantitatively, the relative errors of initial pressure $E_{p_{0}^{i}}$ and speed of sound $E_{c}$ between the reconstructed and true parameter distributions were calculated as
\begin{equation}
\begin{dcases}
       E_{p_{0}^{i}} =  100\% \cdot \frac{|| p_{0,\mathrm{true}}^{i} - p_{0,\mathrm{rec}}^{i} ||}{|| p_{0,\mathrm{true}}^{i} ||}, \\
       E_c =  100\% \cdot \frac{|| c_{\mathrm{true}} - c_{\mathrm{rec}} ||}{|| c_{\mathrm{true}} ||}, 
\end{dcases}
    \label{eq:RelativeError}
\end{equation}
where $p_{0,\mathrm{true}}^{i}$ and $c_{\mathrm{true}}$ are the true, and $p_{0,\mathrm{rec}}^{i}$ and $c_{\mathrm{rec}}$ are the reconstructed parameter distributions for the initial pressure and speed of sound, respectively.

\subsubsection{Results}
\label{sec:SimplePhantom_Results}

\begin{figure}[p]
\centering
    \includegraphics[width=0.85\linewidth]{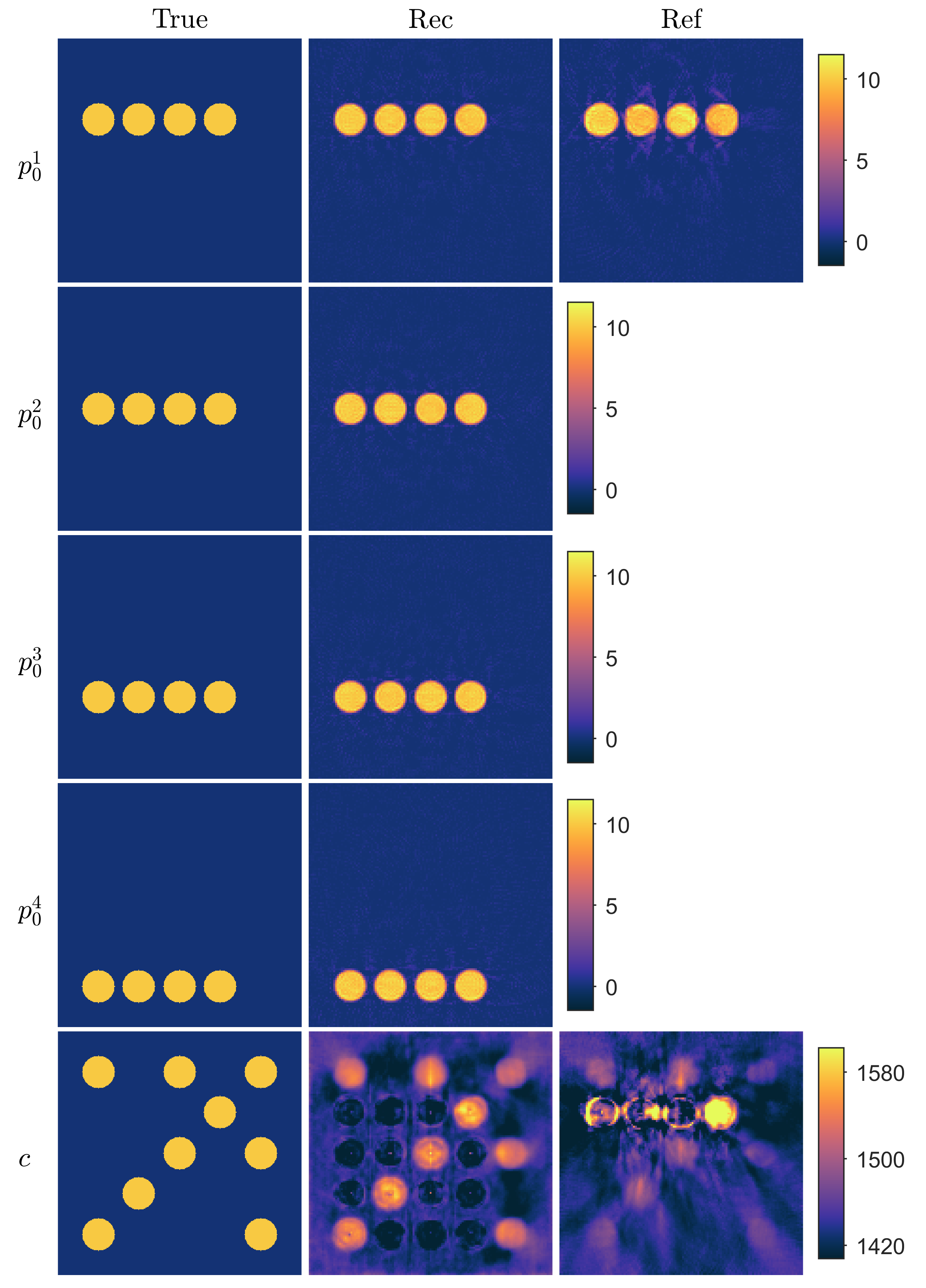}	
    \caption{Reconstructions in a simple geometry phantom with a low speed of sound contrast. First column: True parameters. Second column: Reconstructed initial pressure $p_{0}^{1}$, $p_{0}^{2}$, $p_{0}^{3}$ and $p_{0}^{4}$, and speed of sound $c$ distributions using the proposed approach. Third column: Reconstructed initial pressure $p_0^{\mathrm{1}}$ and speed of sound $c$ distributions using the reference approach. Units are in $\mathrm{Pa}$ and $\mathrm{m/s}$ for initial pressures and speed of sound, respectively.}
    \label{fig:JR_simplecircular}
\end{figure}

Reconstructed initial pressure and speed of sound distributions for the simple phantom with a low speed of sound contrast are shown in Fig. \ref{fig:JR_simplecircular}. As we can see, the reconstructed initial pressure distributions resemble the true parameters well. Locations and parameter values of the inclusions are close to the true parameter. For the reconstructed speed of sound, the shape of the inclusions resembles the true target distributions, but the speed of sound values are a bit lower than the true inclusion values. Also the reconstructed background speed of sound suffers from artefacts, mainly in the vicinity of the initial pressure inclusions. The reconstructions calculated using the reference approach, where a (single) initial pressure and the speed of sound distributions were reconstructed, are also presented in Fig. \ref{fig:JR_simplecircular}. Similarly as with the proposed approach, the reconstructed initial pressure distribution resembles the true parameter in shape and parameter values. However, the reconstructed speed of sound suffers highly from artefacts. The inclusions are hardly recognisable and values are notably smaller than in the true inclusions, except in the case of the inclusion that matches in position with an initial pressure inclusion. The calculated relative errors are presented in Table \ref{tab:Relative_errors_SimpleTarget}. Relative error values for the initial pressure and speed of sound obtained with the proposed approach are lower than when compared to the reference approach.

\begin{table}[tb!]
\renewcommand{\arraystretch}{1.2}
\centering
\caption{Relative errors of reconstructed initial pressures $E_{p_{0}^{1}} \: (\%)$, $E_{p_{0}^{2}} \: (\%)$, $E_{p_{0}^{3}} \: (\%)$ and $E_{p_{0}^{4}} \: (\%)$, and speed of sound $E_{c} \: (\%)$ distributions in a simple geometry phantom.} 
\vspace{1mm}
\begin{tabular}{l c c c c c}
\hline
&   $E_{p_{0}^{1}} $ & $E_{p_{0}^{2}}$ & $E_{p_{0}^{3}}$ & $E_{p_{0}^{4}}$ & $E_{c}$\\ [0.1cm] \hline
 Low contrast & 17.0  & 17.5 & 17.3 & 20.0 & 2.0 \\ 
 Reference & 24.9 &  &  &  & 3.2 \\ \hline 
 Water bath &  17.5 & 18.2 & 17.3 & 18.8 & 2.1  \\ 
 Reference & 26.0 &  &  &  & 3.5 \\ \hline 
 High contrast & 24.8 & 23.9 & 22.1 & 26.0 & 12.0 \\
 Reference & 42.8 &  &  &  & 14.5 \\ \hline 
\end{tabular}
\label{tab:Relative_errors_SimpleTarget}
\end{table}

\begin{figure}[p]
\centering
    \includegraphics[width=0.85\linewidth]{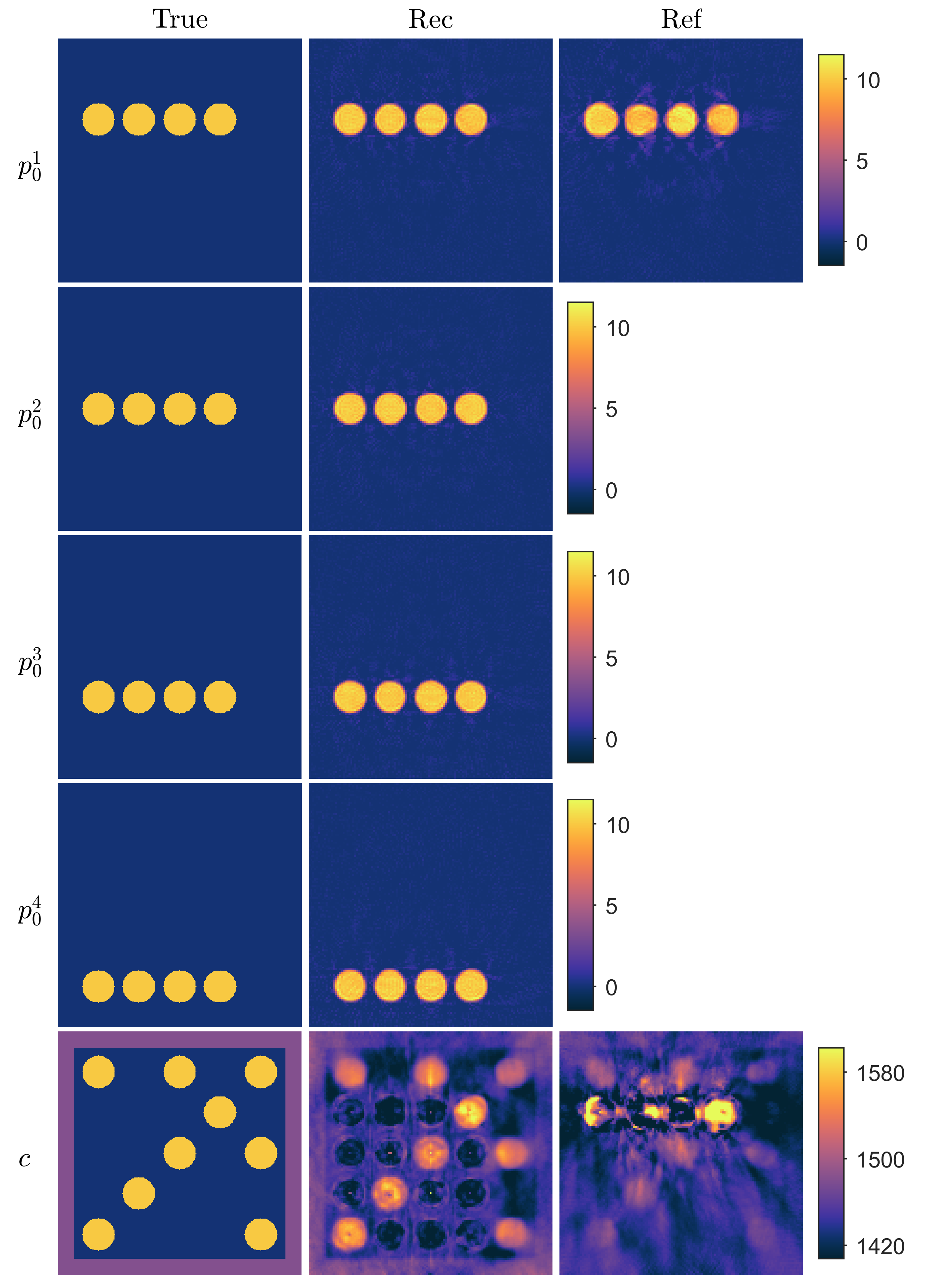}	
    \caption{Reconstructions in a simple geometry phantom with a low speed of sound contrast and a water bath mimicking layer. First column: True parameters. Second column: Reconstructed initial pressure $p_{0}^{1}$, $p_{0}^{2}$, $p_{0}^{3}$ and $p_{0}^{4}$, and speed of sound $c$ distributions using the proposed approach. Third column: Reconstructed initial pressure $p_0^{\mathrm{1}}$ and speed of sound $c$ distributions using the reference approach. Units are in $\mathrm{Pa}$ and $\mathrm{m/s}$ for initial pressures and speed of sound, respectively.}
    \label{fig:JR_simplecircular_water}
\end{figure}

Reconstructed initial pressure and speed of sound distributions for the simple phantom with a low speed of sound contrast and a water bath -mimicking layer are shown in Fig. \ref{fig:JR_simplecircular_water}. As it can be seen, the initial pressure distributions reconstructed with the proposed approach resemble the true parameter distributions well. Also, the reconstructed speed of sound resembles the true parameter as the circular inclusions and the water mimicking layer can be distinguished. However, similarly as in the previous simulation, the speed of sound values are a bit lower and some artefacts are visible. In the case of the reference approach (Fig. \ref{fig:JR_simplecircular_water}), the shape and values of the initial pressure inclusions resembles the true inclusions. The speed of sound distribution, on the other hand, suffers highly from artefacts, and the inclusions are difficult to distinguish. These observations can also be verified quantitatively with the calculated relative errors (see Table. \ref{tab:Relative_errors_SimpleTarget}), where the relative errors are lower with the proposed approach than with the reference approach. 

\begin{figure}[p]
\centering
    \includegraphics[width=0.85\linewidth]{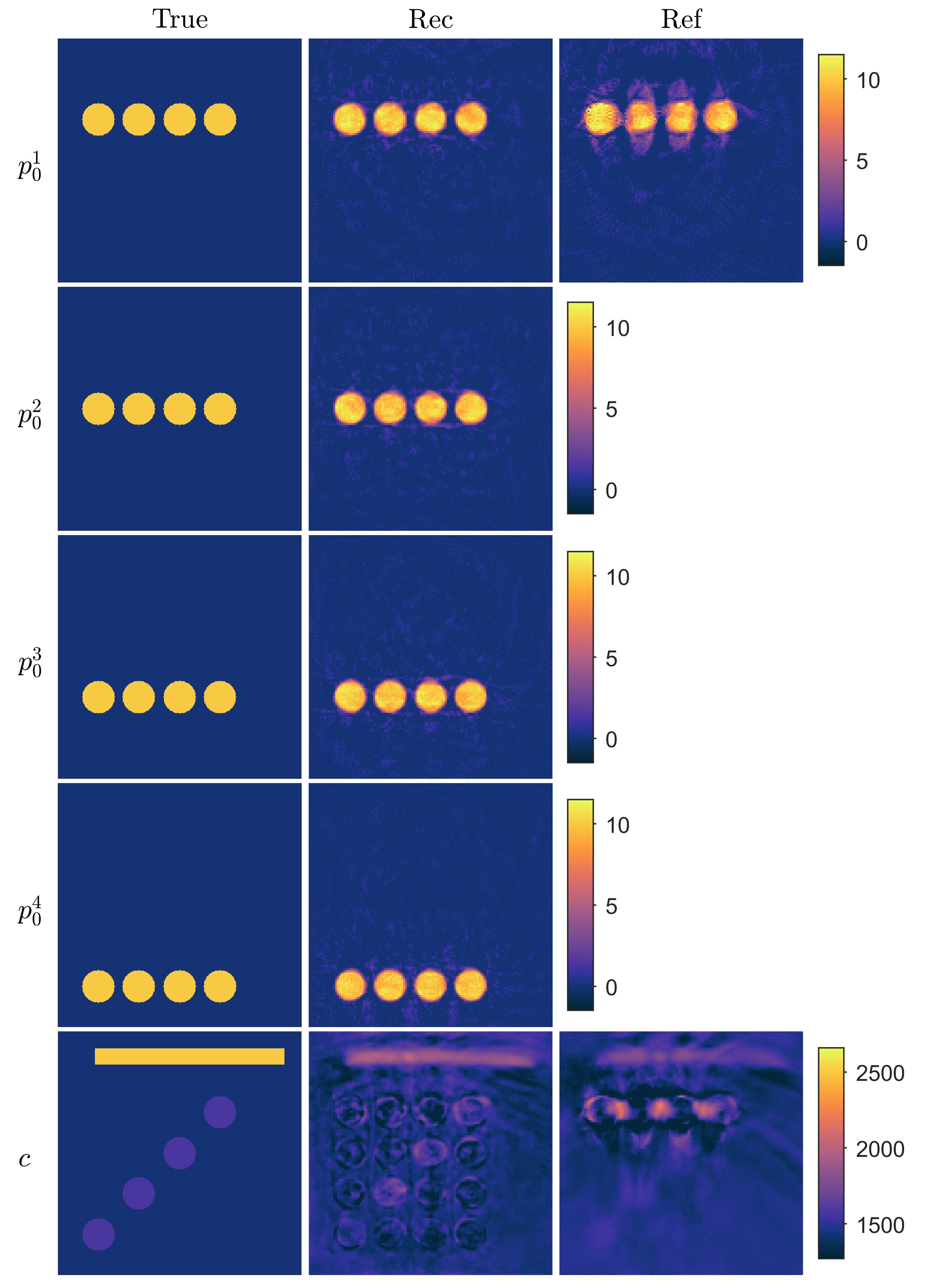}	
    \caption{Reconstructions in a simple geometry phantom with a high speed of sound contrast. First column: True parameters. Second column: Reconstructed initial pressure $p_{0}^{1}$, $p_{0}^{2}$, $p_{0}^{3}$ and $p_{0}^{4}$, and speed of sound $c$ distributions using the proposed approach. Third column: Reconstructed initial pressure $p_0^{\mathrm{1}}$ and speed of sound $c$ distributions using the reference approach. Units are in $\mathrm{Pa}$ and $\mathrm{m/s}$ for initial pressures and speed of sound, respectively.}
    \label{fig:JR_simplecircular_bone}
\end{figure}

Initial pressure and speed of sound reconstructions for the simple phantom and a high speed of sound contrast are shown in Fig. \ref{fig:JR_simplecircular_bone}. Now, the reconstructed initial pressure distributions resemble the true parameter distributions. However, all inclusions show small artefacts that were not present in the previous low-contrast phantom simulations. In the reconstructed speed of sound, the shape of the high contrast inclusion is detectable but the parameter values are lower than the true parameters. In addition, artefacts are visible in the reconstructed image. In the case of the reference approach (Fig. \ref{fig:JR_simplecircular_bone}), the reconstructed initial pressure distribution resembles the true parameters. However, it suffers from larger artefacts than the proposed approach, especially in the vicinity of the inclusions. The speed of sound reconstruction suffers highly from artefacts and the high contrast inclusions are barely distinguishable. Comparing the relative errors (Table. \ref{tab:Relative_errors_SimpleTarget}) supports these findings: the relative errors are  significantly lower with the proposed approach than with the reference approach.

Based on the simulations performed with a simplified numerical phantom, it appears that multiple initial pressure and speed of sound distributions can be reconstructed more accurately than a single initial pressure and speed of sound distributions. It also seems that, when the target contains large variations in the speed of sound values, reconstructing the speed of sound is more difficult than in a low-contrast case. 

\subsection{Tissue-mimicking phantom}
\label{sec:TissuePhantom}

\subsubsection{Phantom and data simulation}
\label{sec:TissuePhantom_TargetAndDatasim}

In the second set of simulations, a more realistic tissue-mimicking photoacoustic phantom was used. Furthermore, modeling of light transport and absorption was included in the data simulation. A numerical breast phantom, that is described in \cite{Park_2023_JBO}, was taken as a target. 2D cross-sectional (coronal plane) slices of the optical parameters and speed of sound distributions were extracted from a 3D numerical phantom and further cropped to match the simulation geometry used in this work  (described in the previous section). The initial pressure distributions were simulated using a light propagation model. As a light propagation model, the diffusion approximation was used \cite{Tarvainen_2012}, see \ref{sec:appendix_DA}. Three different approaches to generate multiple initial pressures in the target were studied: illuminations from different directions, using multiple wave-lengths of light, and introducing additional absorbers on the boundary of the target.

\begin{figure}[tb!]
    \centering
	\includegraphics[width=0.85\linewidth]{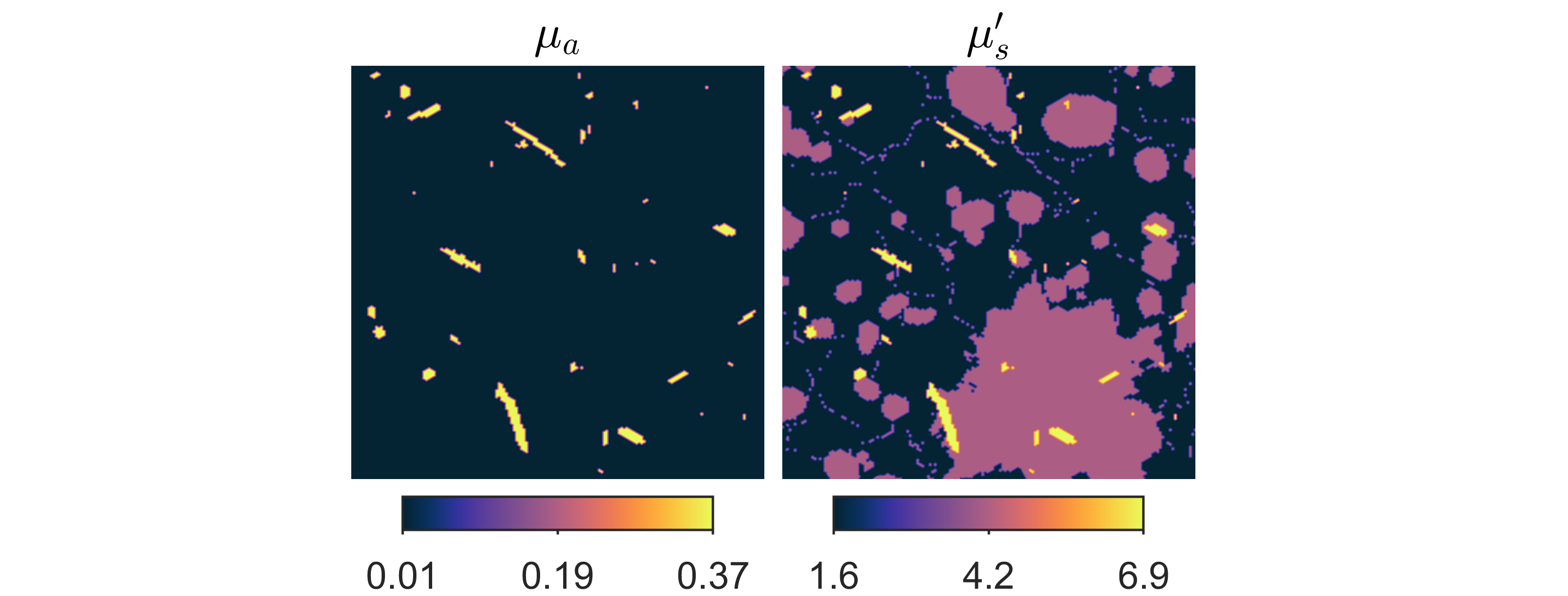}	
	\caption{Optical absorption $\mu_{\mathrm{a}}$ and reduced scattering $\mu_{\mathrm{s}}'$ coefficients used to simulate initial pressure distributions when multiple illuminations were used. Units are in $\mathrm{mm^{-1}}$. }
	\label{fig:Optical_parameters}
\end{figure}

In the first simulation, illuminations from different directions were used to generate different initial pressure distributions in the target. The optical parameters of the target (absorption and reduced scattering) are illustrated in Fig. \ref{fig:Optical_parameters}. The target was illuminated from different directions using a planar illumination with the width of the side of the phantom, leading to four different initial pressure distributions: $p_{0}^{1}$ (illumination from the top), $p_{0}^{2}$ (illumination from the left), $p_{0}^{3}$ (illumination from the bottom) and $p_{0}^{4}$ (illumination from the right). The simulated initial pressure and speed of sound distributions are presented in Fig. \ref{fig:MultiIllumination} that is shown later in Sec. \ref{sec:TissuePhantom_Results}. As a reference, a single illumination from the top and thus initial pressure $p_{0}^{1}$ was used.

\begin{figure}[tb!]
    \centering
	\includegraphics[width=0.85\linewidth]{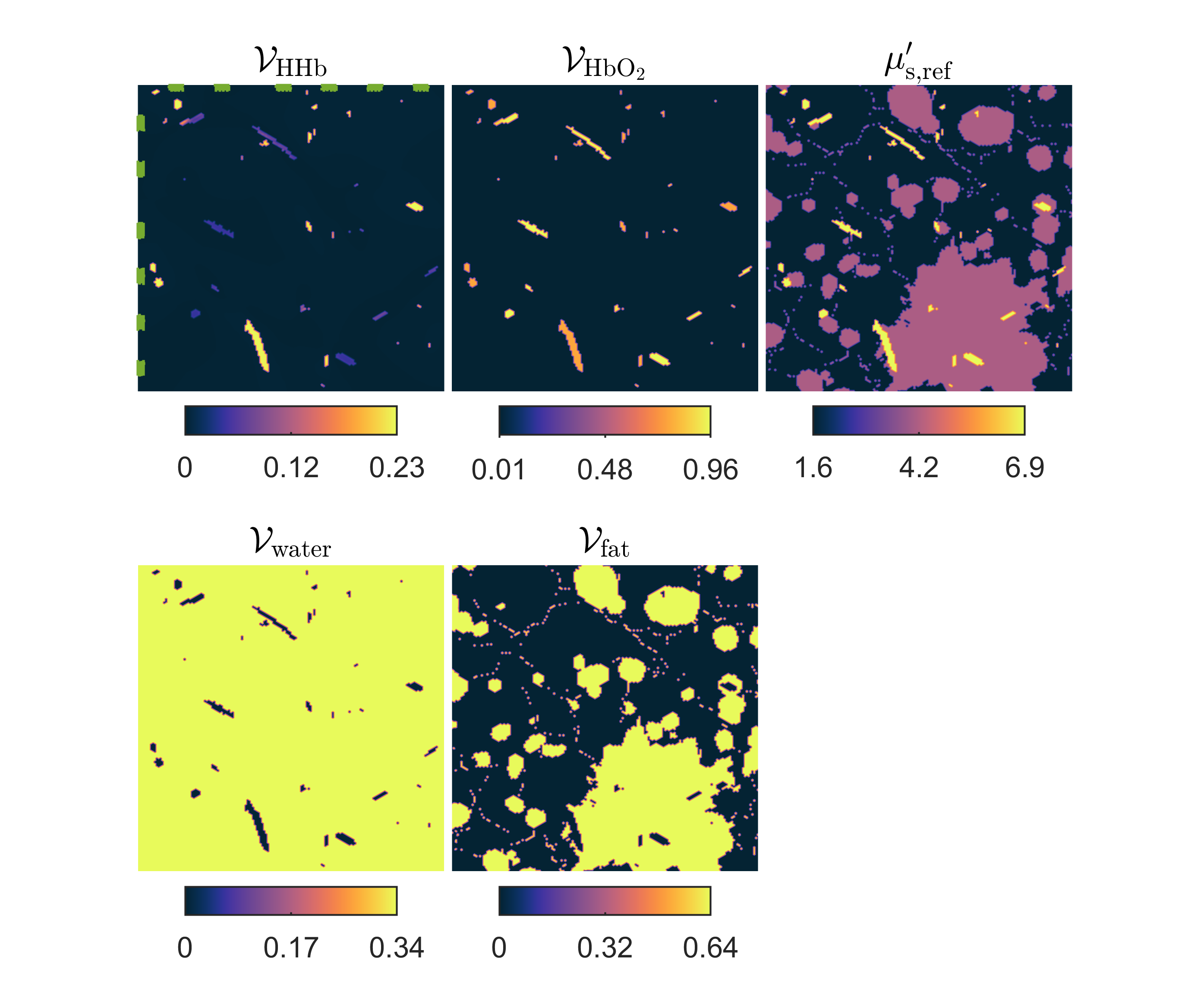}	
	\caption{Volume fractions of deoxygenated $\mathcal{V}_{\mathrm{HHb}}$, and oxygenated $\mathcal{V}_{\mathrm{HbO_2}}$ hemoglobin, water $\mathcal{V}_{\mathrm{water}}$ and fat $\mathcal{V}_{\mathrm{fat}}$, and reference scattering $\mu_{\mathrm{s,ref}}'$ at wavelength $\lambda_{\mathrm{ref}} = 800 \: \mathrm{nm}$ used to simulate initial pressure distributions when multiple wavelengths or exogenous absorbers were utilised. Locations of exogenous absorbers are highlighted with green color in the top left figure. Units are in $\mathrm{mm^{-1}}$ for reference scattering. }
	\label{fig:Spectral_parameters}
\end{figure}

In the second simulation, we studied a situation where the target cannot be illuminated from all directions, and instead the initial pressure distributions were simulated by using multiple wavelengths of light. In this case, the target was assumed to consist of oxygenated and deoxygenated hemoglobin, water and fat. Target optical (spectral) parameters are illustrated in Fig. \ref{fig:Spectral_parameters}. Four different initial pressure distributions $p_{0}^{1}$, $p_{0}^{2}$, $p_{0}^{3}$ and $p_{0}^{4}$ were simulated by using wavelengths $650 \: \mathrm{nm}, 750 \: \mathrm{nm},850 \: \mathrm{nm} \; \mathrm{and} \; 950 \: \mathrm{nm}$, respectively. In all illuminations, a planar illumination that covered the top and left sides of the domain was used. The simulated initial pressure and speed of sound distributions shown in Fig. \ref{fig:MultiWavelength} that is shown later in Sec. \ref{sec:TissuePhantom_Results}. For the reference approach, the initial pressure $p_{0}^{1}$ was used, i.e. data simulated with the wavelength $\lambda = 650 \: \mathrm{nm}$.

In the third simulation, photoacoustic datasets were simulated by including additional absorbers to the imaged target. Now, the same multi-wavelength optical phantom was used as in the previous simulations. However, additional absorbers were added on the boundary. One initial pressure distribution $p_{0}^{1}$ was simulated without adding any exogenous absorbers and three initial pressure distributions $p_{0}^{2}$, $p_{0}^{3}$ and $p_{0}^{4}$ were simulated by adding four exogenous absorbers with an absorption coefficient value of $\mu_{\mathrm{a}} = 0.3 \: \mathrm{mm^{-1}}$ to different positions in the target (see Fig. \ref{fig:Spectral_parameters}).  The simulated initial pressure and speed of sound distributions shown in Fig. \ref{fig:Absorbers} (Sec. \ref{sec:TissuePhantom_Results}). In the reference approach, the initial pressure target was the one without an exogenous absorber i.e. $p_{0}^{1}$.

Photoacoustic data generated by the initial pressure distributions was simulated with the same discretisation and parameter settings as the data of the simple phantom (see Table \ref{tab:GridParameters}). Furthermore, additive Gaussian distributed noise with $1\%$ of peak-to-peak amplitude of the corresponding simulated photoacoustic data, was added to the data.

\subsubsection{Image reconstruction}
\label{sec:TissuePhantom_ImageRec}

Image reconstruction was performed using a multigrid approach, similarly as with simple phantom, (see Secs. \ref{sec:multigrids} and \ref{sec:SimplePhantom_ImageRec}). The tissue-mimicking target had structure with higher resolution compared to the simple target, and thus higher discretisation in the image reconstruction was needed. Details of the multigrid discretisation are presented in Table \ref{tab:GridParameters}. 

As a prior model, Ornstein-Uhlenbeck prior was used. Prior mean $\eta_z$ for each parameter was chosen as the median value of interval of the true parameter distribution, and standard deviations $\sigma_z$ were chosen so that $\sigma_z = 1/4\left( \max(z_{\mathrm{true}}) - \min(z_{\mathrm{true}})  \right)$. This means that $95 \%$ of the parameter values were expected to lie within the values of the true parameter range. Value $\tau = 1 \: \mathrm{mm}$ for characteristic length scale was used. Similarly as in the simple target simulations, noise model of 1\% additive and zero mean noise was used for every dataset corresponding to different initial pressures

Similarly as in simple target simulations, initial guess $x_0$ for the reconstructed parameters was chosen such that the initial pressure was zero and speed of sound was the mean of the prior. In addition, bound constraints $L_{p_0} = 0$ with no upper bound for initial pressure, and $L_c = 1300 \: \mathrm{m/s}$ and $U_c= 1800 \: \mathrm{m/s}$ for speed of sound were used.


\subsubsection{Results}
\label{sec:TissuePhantom_Results}

\begin{figure}[p]
\centering
    \includegraphics[width=0.85\linewidth]{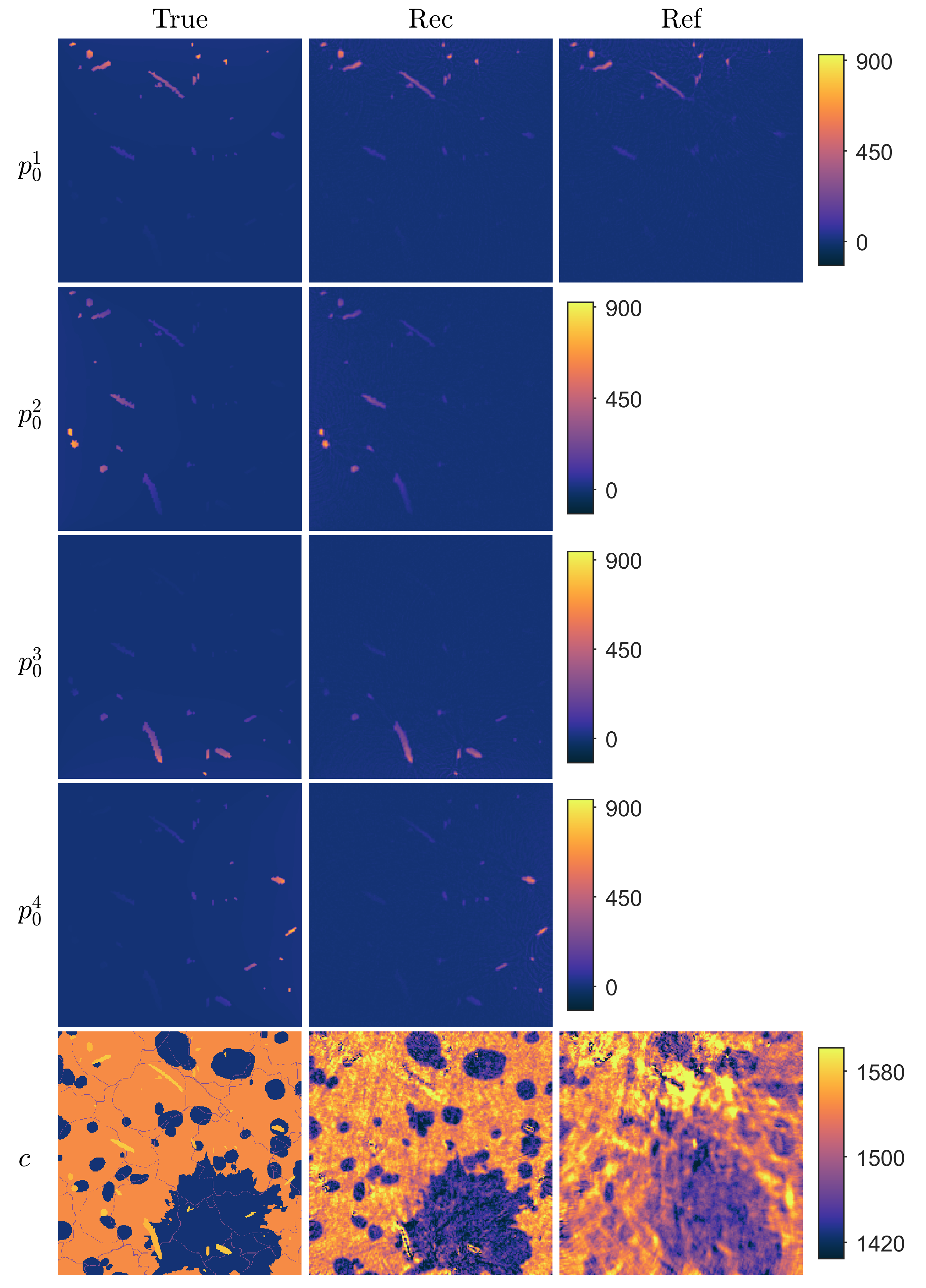}	
    \caption{Reconstructions in a tissue-mimicking target with multiple illuminations. First column: True parameters. Second column:  Reconstructed initial pressure $p_{0}^{1}$, $p_{0}^{2}$, $p_{0}^{3}$ and $p_{0}^{4}$, and speed of sound $c$ distributions using the proposed approach. Third column: Reconstructed initial pressure $p_0^{\mathrm{1}}$ and speed of sound $c$ distributions using the reference approach. Units are in $\mathrm{Pa}$ and $\mathrm{m/s}$ for initial pressures and speed of sound, respectively.}
    \label{fig:MultiIllumination}
\end{figure}

Reconstructed distributions of the initial pressure and speed of sound, when the true initial pressures were simulated by planar illuminations from different directions, are shown in Fig. \ref{fig:MultiIllumination}. As it can be seen, the reconstructed initial pressure distributions resemble closely the true parameter distributions. In the reconstructed speed of sound distribution, the shape and location of the heterogeneities are well distinguishable and the speed of sound values are close to the true speed of sound. Fig. \ref{fig:MultiIllumination} also shows the reference reconstructions of the initial pressure and speed of sound, where the true initial pressure distribution was simulated with a single illumination from the top. In this situation, the initial pressure distribution resembles the true parameter distribution. However, the speed of sound distribution is highly distorted and the target structure cannot be accurately distinguished. The calculated relative errors are presented in Table \ref{tab:Relative_errors_TissueTarget}. These support the visual findings as the relative errors of the initial pressure $p_{0}^{1}$ and speed of sound are lower for proposed approach than of the reference. Thus, it appears that using multiple illuminations from different directions could be used to improve simultaneous reconstruction of initial pressure and speed of sound.

\begin{table}[tb!]
\renewcommand{\arraystretch}{1.2}
\centering
\caption{Relative errors of reconstructed initial pressures $E_{p_{0}^{1}} \: (\%)$, $E_{p_{0}^{2}} \: (\%)$, $E_{p_{0}^{3}} \: (\%)$ and $E_{p_{0}^{4}} \: (\%)$, and speed of sound $E_{c} \: (\%)$ distributions in a tissue-mimicking phantom.} 
\vspace{1mm}
\begin{tabular}{l c c c c c}
\hline
 &   $E_{p_{0}^{1}}$ & $E_{p_{0}^{2}}$ & $E_{p_{0}^{3}}$ & $E_{p_{0}^{4}}$ & $E_{c}$ \\ [0.1cm] \hline
 Multi-illumination & 28.1 & 26.9 & 28.6 & 30.6 & 1.7 \\ 
 Reference & 38.4 &  &  &  & 3.1 \\  \hline
 Multi-wavelength & 49.0 & 48.3 & 48.7 & 48.8 & 5.6 \\ 
 Reference & 46.7 &  &  &  & 4.3 \\ \hline
 Exogeneous absorbers & 29.8 & 27.1 & 28.1 & 28.2 & 2.2 \\
 Reference & 46.7 &  &  &  & 4.3 \\ \hline
\end{tabular}
\label{tab:Relative_errors_TissueTarget}
\end{table}

\begin{figure}[p]
\centering
    \includegraphics[width=0.85\linewidth]{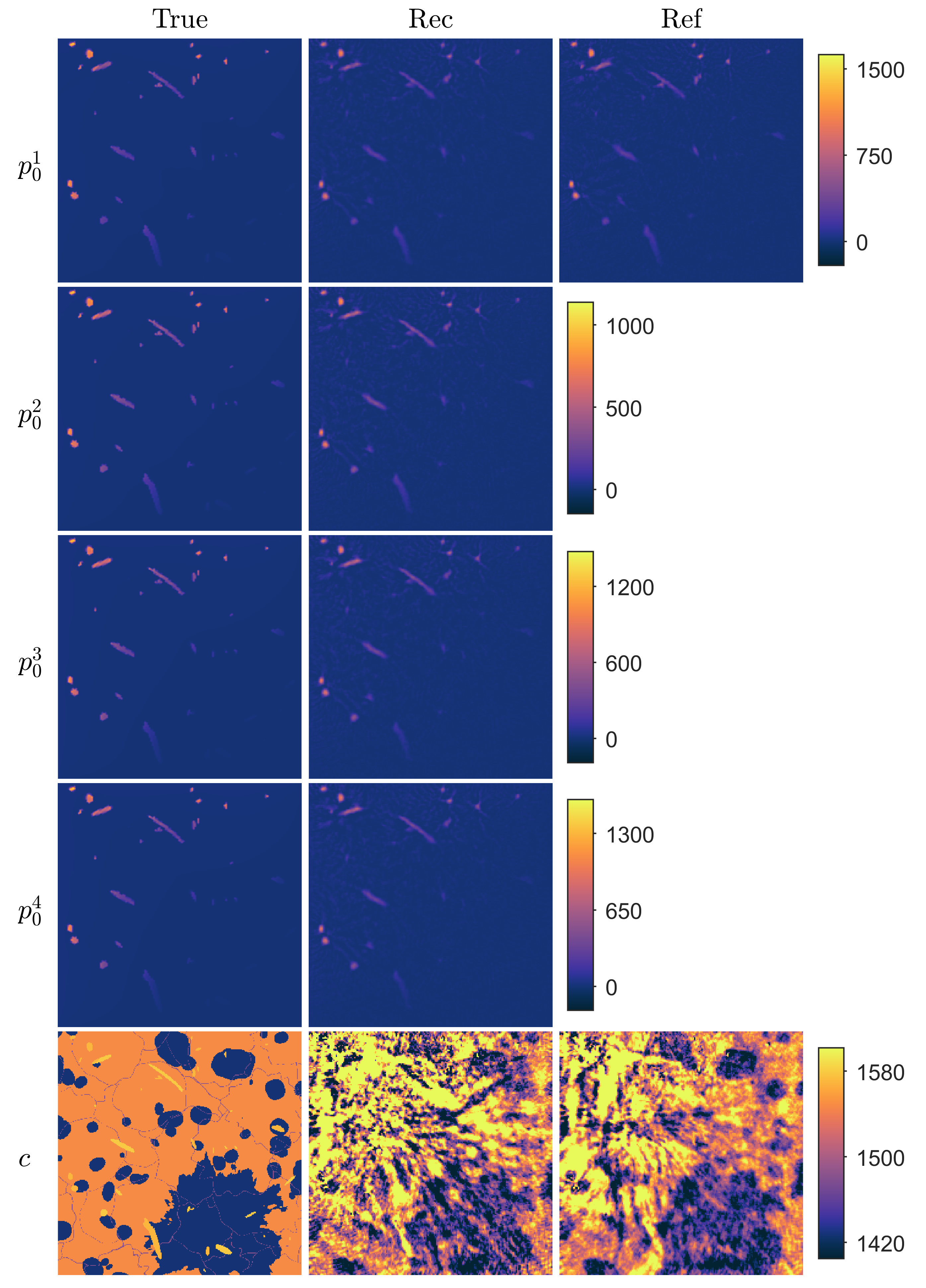}	
    \caption{ Reconstructions in a tissue-mimicking target with multiple wavelengths of light. First column: True parameters. Second column:  Reconstructed initial pressure $p_{0}^{1}$, $p_{0}^{2}$, $p_{0}^{3}$ and $p_{0}^{4}$, and speed of sound $c$ distributions using the proposed approach. Third column: Reconstructed initial pressure $p_0^{\mathrm{1}}$ and speed of sound $c$ distributions using the reference approach. Units are in $\mathrm{Pa}$ and $\mathrm{m/s}$ for initial pressures and speed of sound, respectively.}
    \label{fig:MultiWavelength}
\end{figure}

Reconstructed initial pressure and speed of sound distributions, in the case where multiple wavelengths of light were utilised to simulate different initial pressure distributions, are shown in Fig. \ref{fig:MultiWavelength}. The reconstructed initial pressure distributions resemble the true parameters. However, small amount of artefacts are visible in the background. The reconstructed speed of sound distribution is distorted and suffers from artefacts. The reference reconstructions, when a single wavelength of light is utilised, are also presented in Fig. \ref{fig:MultiWavelength}. Similarly as in the case of multiple wavelengths, the initial pressure distribution resembles the true parameter distribution and only small artefacts are visible in the reconstruction. However, the reconstructed speed of sound is distorted and suffers from artefacts. Comparing the relative errors, presented in Table \ref{tab:Relative_errors_TissueTarget}, reveals that the relative errors of the reference approach are lower than the proposed approach. Thus, it seems that utilising multiple wavelengths is not necessary enough to improve the simultaneous reconstruction of initial pressure and speed of sound. We believe that this is related to whether absorption of light at different wavelengths by different chromophores produces spatially different initial pressure distributions. Now, in this simulation, the initial pressure distributions (Fig. \ref{fig:MultiWavelength}), are spatially very similar, and may not ease the ill-posedness of the inverse problem.

\begin{figure}[p]
\centering
    \includegraphics[width=0.85\linewidth]{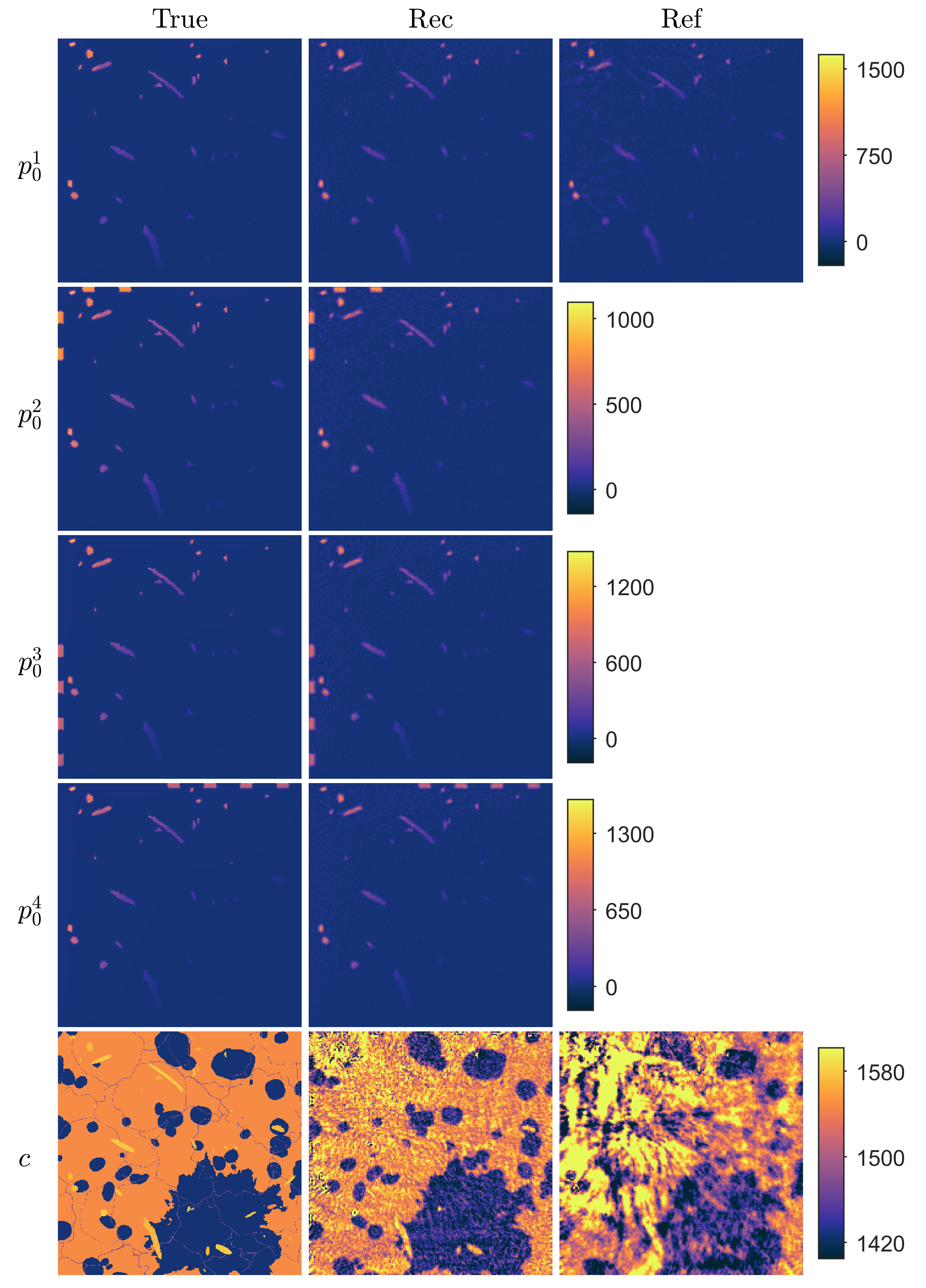}	
    \caption{Reconstructions in a tissue-mimicking target with exogenous absorbers. First column: True parameters. Second column:  Reconstructed initial pressure $p_{0}^{1}$, $p_{0}^{2}$, $p_{0}^{3}$ and $p_{0}^{4}$, and speed of sound $c$ distributions using the proposed approach. Third column: Reconstructed initial pressure $p_0^{\mathrm{1}}$ and speed of sound $c$ distributions using the reference approach. Units are in $\mathrm{Pa}$ and $\mathrm{m/s}$ for initial pressures and speed of sound, respectively.}
    \label{fig:Absorbers}
\end{figure}

Reconstructed initial pressure and speed of sound distributions in the case where exogenous absorbers were placed on the boundary of the imaged are shown in Fig. \ref{fig:Absorbers}. Now, the reconstructed initial pressure distributions resemble the true parameters well. The reconstructed speed of sound distribution is similar when compared to the true distribution since the shape, location and values of the heterogeneities are well distinguished. However, the image suffers from some artefacts. The reconstructions with the reference approach, i.e. without the additional absorbers, are also shown in Fig. \ref{fig:Absorbers}. Also in this case, the initial pressure distribution resembles the true parameter distribution. However, it has more artefacts when compared to the reconstruction with the proposed approach. On the other hand, the reconstructed speed of sound suffers from severe artefacts. The relative errors given in Table \ref{tab:Relative_errors_TissueTarget} support the visual findings, as the relative errors of the initial pressure and speed of sound are lower with the proposed approach than the values calculated with the reference approach. Thus, it seems that utilising exogenous absorbers could aid in simultaneous reconstruction of initial pressure and speed of sound.

\section{Discussion and Conclusions}
\label{sec:discussion}

In this work, simultaneous reconstruction of initial pressure and the speed of sound distributions in PAT was studied. An approach were multiple initial pressure distributions are utilised to alleviate ill-posedness of the image reconstruction problem was proposed. The approach was compared against a (conventional) approach where a single initial pressure distribution and speed of sound distributions are reconstructed. 

The approach was evaluated with numerical simulations. The results show that multiple initial pressure distributions and the speed of sound distribution can be reconstructed simultaneously from photoacoustic data. Furthermore, the quality of the reconstructions was found to be higher compared to a reference approach of reconstructing a single initial pressure distribution and speed of sound simultaneously. Furthermore, results show that in more realistic imaging situations, these different initial pressures used could be formed by illuminating the target from multiple directions or by placing additional optical absorbers to the target. However, utilising only multiple wavelengths of light without additional absorbers, may not generate such initial pressure distributions that they would be different enough to ease the ill-posedness of the image reconstruction problem. 

Overall, the problem of reconstructing the initial pressure and speed of sound simultaneously is challenging and advanced methods to solve this problem are needed. In addition to requirement of careful implementation of these methods, such as including bound constraints and the use of multigrid methods, this also increases the computational expensive nature of the problem by the need of, for example, fine discretisations and multiple iterations. The future steps for research include, for example, extension of the work in realistic 3D geometries, and experiments with real photoacoustic data.

\appendix
\section*{Appendix A. Simulating the initial pressure distributions with the diffusion approximation}
\
\label{sec:appendix_DA}
\setcounter{table}{0}

In this work, the diffusion approximation (DA) was used to simulate absorbed optical energy density leading to different initial pressure distributions in the target. The DA is of the form
\begin{equation}
\begin{cases}
    &-\nabla \cdot \kappa(r,\lambda) \nabla \Phi(r,\lambda) + \mu_{\mathrm{a}}(r,\lambda) \Phi(r,\lambda) = 0, \quad r\in \Omega \\ 
    &\Phi(r,\lambda) + \frac{1}{2\varsigma_d} \kappa(r,\lambda) \nabla \Phi(r,\lambda) \cdot \hat{n} = 
    \begin{cases}
    \frac{s(r,\lambda)}{\varsigma_d}, \quad &r \in \epsilon \\
    0, \quad &r \in \epsilon \setminus \partial \Omega
    \end{cases}
\end{cases}
    \label{eq:DA}
\end{equation}
where $\Phi(r,\lambda)$ is the photon fluence at wavelength $\lambda$ and $\kappa(r,\lambda) = \left(d(\mu_{\mathrm{a}}(r,\lambda) + \mu_{\mathrm{s}}'(r,\lambda)\right)^{-1}$ is the diffusion coefficient where $\mu_{\mathrm{a}}(r,\lambda)$ and $\mu_{\mathrm{s}}'(r,\lambda)$ are absorption and reduced scattering coefficients \cite{Arridge_1999_IP, Ishimaru_1978}. Further, $s(r,\lambda)$ is the light source at a source position $\epsilon \subset \partial \Omega$, $\varsigma_d$ is the dimension dependant parameter ($\varsigma_2 = \frac{1}{\pi}$ in 2D) and $\hat{n}$ is an outward unit normal. Wavelength dependent optical absorption and reduced scattering coefficients can be calculated as
\begin{align}
\begin{dcases}
    \mu_{\mathrm{a}}(r,\lambda) = \sum_{l=1}^{L} \mathcal{V}_{l}(r)\mu_{\mathrm{a},l}(\lambda), \\\mu_{\mathrm{s}}'(r,\lambda) = \mu_{\mathrm{s,ref}}'(r)\left(\frac{\lambda}{\lambda_{\mathrm{ref}}} \right),
\end{dcases}
\end{align}
where $\mathcal{V}_{l}$,  $l = 1,...,L$ are the volume fractions of $L$ chromophores and $\mu_{\mathrm{a},l}$ are the absorption spectra value of each distinctive chromophore \cite{Jacques_2013}. Absorption spectra values for different chromophores, that are used in this work, are presented in Table \ref{tab:Absorption_Spectra}. Further, $\mu_{\mathrm{s,ref}}'(r)$ is the reduced scattering coefficient at reference wavelength $\lambda_{\mathrm{ref}}$. The initial pressure can be calculated from photon fluence as
\begin{align}
    p_0(r,\lambda) = \gamma(r)\mu_{\mathrm{a}}(r,\lambda)\Phi(r,\lambda),
\end{align}
where $\gamma(r)$ is the Gr\"{u}neisen parameter that describes the photoacoustic efficiency of the target. In this work, the Gr\"{u}neisen parameter is modelled as constant $\gamma(r) = 1$.

Solution of the diffusion approximation was numerically approximated using the finite element method (FEM) \cite{Tarvainen_2012}. For this, the target was discretised into a triangular grid consisting of $47310$ elements and $23942$ nodes, where the optical parameters and the calculated absorbed optical energy were presented in a piece-wise linear basis. Simulated initial pressures in the triangular FEM-grid were linearly interpolated to a pixel grid that was used to simulate the photoacoustic data.

\begin{table}[tb!]
\renewcommand{\arraystretch}{1.2}
\centering
\caption{Absorption spectra values used for pure oxygenated $\mu_{\mathrm{a,oxy}} \: (\mathrm{mm^{-1}})$ and deoxygenated $\mu_{\mathrm{a,deoxy}} \: (\mathrm{mm^{-1}})$ hemoglobin, water $\mu_{\mathrm{a,water}} \: (\mathrm{m^{-1}})$ and fat $\mu_{\mathrm{a,fat}} \: (\mathrm{m^{-1}})$ with different wavelengths of light $\lambda  \: (\mathrm{nm})$ \cite{Jacques_2013,Hale_1973, vanVeen_2005}.}
\vspace{0.2cm}
\begin{tabular}{c c c c c}
\hline
 $\lambda$ & $\mu_{\mathrm{a,oxy}}$ & $\mu_{\mathrm{a,deoxy}}$ & $\mu_{\mathrm{a,water}}$ & $\mu_{\mathrm{a,fat}}$ \\ \hline
 650 & 0.17 & 1.74 & 0.32 & 0.47 \\
 750 & 0.25 & 0.69 & 2.6 & 0.97 \\
 850 & 0.52 & 0.35 & 2.0 & 0.64 \\
 950 & 0.60 & 0.32 & 39.0 & 3.7 \\ \hline
\end{tabular}
\label{tab:Absorption_Spectra}
\end{table}

\subsection*{Disclosures}

The authors declare no conflicts of interest.

\subsection* {Code, Data, and Materials Availability} 

The unrestricted data of this study are available upon reasonable request from the corresponding author.

\subsection* {Acknowledgments}

This project was supported by the European Research Council (ERC) under the European Union’s Horizon 2020 research and innovation programme (grant agreement No 101001417 - QUANTOM) and by the Research Council of Finland (Centre of Excellence in Inverse Modelling and Imaging grant 353086, Flagship of Advanced Mathematics for Sensing Imaging and Modelling grant 358944), Finnish Cultural Foundation (South Savo Regional Fund), and Magnus Ehrnrooth Foundation. B.C. acknowledges the support of the Engineering and Physical Sciences Research Council, UK (EP/W029324/1, EP/T014369/1).

\bibliographystyle{abbrv}

\end{document}